\documentstyle[amsfonts,epsfig,12pt]{article}

\def\mypagenumber{1}

\def\myend{\end{document}}

\normalsize

\newcounter{sxn}

\newcounter{axn}

\date{}

\newdimen\mybaselineskip
\newcommand{\beeq}{\begin{equation}}
\newcommand{\eneq}{\end{equation}}
\newcommand{\be}{\begin{eqnarray}}
\newcommand{\ee}{\end{eqnarray}}
\newcommand{\bpic}{\begin{picture}}
\newcommand{\epic}{\end{picture}}

\def\la{\raise.16ex\hbox{$\langle$} \, }
\def\ra{\, \raise.16ex\hbox{$\rangle$} }

\def\psibar{ \psi \kern-.65em\raise.6em\hbox{$-$} }
\def\mbar{ m \kern-.78em\raise.4em\hbox{$-$}\lower.4em\hbox{} }

\def\n@space{\nulldelimiterspace=0pt \mathsurround=0pt }
\def\huge#1{{\hbox{$\left#1\vbox to 20.5pt{}\right.\n@space$}}}

\def\myskip{\noalign{\kern 8pt}}
\def\myeqspace{\noalign{\kern 10pt}}

\def\boxit#1{$\vcenter{\hrule\hbox{\vrule\kern3pt
    \vbox{\kern3pt\hbox{#1}\kern3pt}\kern3pt\vrule}\hrule}$}
\def\bigbox#1{$\vcenter{\hrule\hbox{\vrule\kern5pt
     \vbox{\kern5pt\hbox{#1}\kern5pt}\kern5pt\vrule}\hrule}$}

\def\ignore#1{{}}

\begin{document}

\bibliographystyle{unsrt}
\footskip 1.0cm

\thispagestyle{empty}
\setcounter{page}{\mypagenumber}

             
\begin{flushright}{
OUTP-01-31-P\\}

\end{flushright}

\vspace{2.5cm}
\begin{center}
{\LARGE \bf {Higher Dimensional Operators or Large Extra Dimensions? }}\\ 
\vskip 1 cm
{\large{Bayram Tekin}}$^{a,}$\footnote{e-mail:~
tekin@thphys.ox.ac.uk}\\
\vspace{.5cm}
$^a${\it Theoretical Physics, University of Oxford, 1 Keble Road, Oxford,
OX1 3NP, UK}\\

\end{center}

\vspace*{2.5cm}


\begin{abstract}
\baselineskip=18pt

We deform gravity with higher curvature terms in four dimensions and 
argue that the non-relativistic limit is of the same form as the 
 non-relativistic limit
of the theories with large extra dimensions. Therefore the experiments
that perform sub-millimeter tests of inverse-square law cannot
distinguish the effects of large extra dimensions from the
effects of higher dimensional operators. In other words instead of
detecting the presence of sub-millimeter dimensions; 
the experiments could be detecting the existence of massive modes of 
gravity with large masses ($\ge 10^{-3}$ eV ). 
\end{abstract}
\vfill

 
\newpage



\normalsize
\baselineskip=22pt plus 1pt minus 1pt
\parindent=25pt

In the context of ``brane world'' pictures 
modification of gravity both at small and large 
distances has been a subject of recent discussions \cite{arkani,anton,randall,ross,
gregory,dvali}. During these discussions an old puzzle, 
about the mass of the gravitons re-emerged: 
van Dam-Veltman-Zakharov discontinuity.

In the papers \cite{veltman,zakharov} it was argued that
predictions of the gravity theory with strictly massless 
gravitons and the theory with arbitrarily small massive gravitons are
different. The story is well-summarized in a recent paper  
\cite{deffayet} but we shall briefly recapture what the issue is
before we address our problem.

The point raised in \cite{veltman,zakharov} for massive gravity
is rather puzzling since it seems that massless limit of massive
gravity is not General Relativity.
In fact starting with the Einstein-Hilbert action
\be
S= {1\over 16\pi G } \int d^4 x\,\sqrt{-g} R, 
\ee
in the weak field limit one obtains the usual Newton's potential
\be
V(r)= -{GM\over r}
\ee
On the other hand if the action is augmented by a Pauli-Fierz
mass term~\footnote{Boulware and Deser \cite{deser} argue that
there is no consistent quantum theory of massive spin-2 
fields in four dimensions.}
 
\be
S= {1\over 16\pi G }  \int d^4 x\,\sqrt{-g} \left \{ R + {m_g^2\over 4}
[h_{\mu\nu}^2 - (h_\mu^\nu)^2] \right \} 
\ee
where $h_{\mu\nu}$ is a small perturbation around the Minkowski space,
one obtains the potential 
\be
V(r)= -{4\over 3} {GM\over r}e^{-m_g r}
\ee
Therefore in the limit of vanishing graviton mass the two potentials 
do not match. One can cure the puzzle by redefining the
Newton's constant of the latter theory but the prediction of massive
gravity for the bending of light around the Sun would be off 
by 25\%. At this level of perturbation theory there seems to be no room for
a massive graviton no matter how small mass it has.

Resolution of this puzzle was  given by Vainshtein \cite{vainshtein,deffayet}
who showed that one needs to be careful
in taking the small $m_g$ limit. Starting from the Schwarzschild solution
in the massive case he showed that $m_g \rightarrow 0$ limit is singular and 
a proper limiting procedure shows that there is in fact no discontinuity.
Another resolution of this puzzle was 
given in the recent works \cite{kogan,porrati} 
who showed that there is no discontinuity
in massless graviton limit if there is a cosmological constant of both signs.
See also the earlier work \cite{higuchi} on $dS_4$. It was
reported in \cite{talk} that there is no discontinuity for generic 
backgrounds with non-constant curvature invariants.

Even after one resolves van-Dam-Veltman-Zakharov puzzle `a la Vainshtein,
clearly the mass one can give to the graviton through Pauli-Fierz term
should be tiny. Experimental astrophysical limits allow 
$m_g$ to be at most  $(10^{25} cm )^{-1}$ \cite{groom}.

There is an other way to study the modifications of gravity in four 
dimensions; namely by adding higher curvature terms in the action.
In this Letter we start with the following action,  
\be
S= \int d^4 x\,\sqrt{-g}\left \{ {R\over 16\pi G } + \alpha R^2
+\beta R_{\mu\nu}R^{\mu\nu}+ \gamma R_{\mu\nu\lambda\delta}
R^{\mu\nu\lambda\delta} \right \}
\label{action1}
\ee
We shall try to answer the following questions.
What are the experimentally allowed values of $\alpha$, $\beta$ and $\gamma$ ?
Do they have to be strictly zero? If not zero how large can they be?  
We will find that the answers to these questions seriously effect the
interpretations of the sub-millimeter experiments on gravity which
try to ``see'' large extra dimensions.~\footnote{I have received arguments
suggesting that working with only quadratic terms is an abuse of the `effective
theory' and one should consider all higher order terms like $R^n$. This objection
is simply wrong for two reasons first of all, the rest of the higher order terms will come into
the picture {\it{only}} at very high energies, namely around the Planck scale. Secondly 
I am considering the propagators for which higher order terms do not contribute.
In any case at the sub-millimeter scale one can disregard all higher order terms
but the ones I consider here.}   
 
Observe that at this stage if one sets
\be
\beta = -4\alpha \hskip 2 cm \gamma = \alpha
\label{topological}
\ee
Then one obtains Einstein-Hilbert-Gauss-Bonnet action
\be
S= \int d^4 x\,\sqrt{-g}\left \{ {R\over 16\pi G } + \alpha ( R^2
-4 R_{\mu\nu}R^{\mu\nu}+ R_{\mu\nu\lambda\delta} 
R^{\mu\nu\lambda\delta} ) \right \}
\ee
This action appears in the string-generated models of gravity 
\cite{boulware,zwiebach}.
The last three terms combine to make a topological invariant 
$\epsilon^{\mu\nu \rho\sigma} \epsilon_{\alpha\beta \gamma\delta}
R_{\mu \nu}\,^{\alpha \beta} R_{\rho \sigma}\,^{\gamma \delta}$
and therefore do not effect the equations of motion in four dimensions. 
In particular Schwarzschild 
solution remains intact and the Newton's potential outside the horizon
of a massive spinless particle is
\be
V(r)= -{G M\over r}
\ee
Clearly it is not possible to detect/measure the
existence of the topological term and $\alpha$ can take any, 
in particular large, value. This Gauss-Bonnet limit 
(\ref{topological}) is not particularly interesting as far as
classical dynamics is concerned.

Now we shall take a different route and start with generic
values of $\alpha$, $\beta$ and $\gamma$ and compute the observable
effects of these dimensionless parameters.

We shall consider the scattering of two massive spin-0 particles 
in the non-relativistic limit.
\be
S_M=  {1\over 2}\int d^4x\,\sqrt{-g}\left \{ g^{\mu\nu} \partial_{\mu}\phi
 \partial_{\nu}\phi - M^2 \phi^2 \right \}
\ee

With no effort one can see that 
the action (\ref{action1}) can be written in the following form
\be
S = \int d^4 x\,\sqrt{-g}\left \{ {R\over 16\pi G } + {\tilde{\alpha}} R^2
+{\tilde{\beta}} R_{\mu\nu}R^{\mu\nu}
 \right \} + S_{GB}
\label{action2} 
\ee
where the Gauss-Bonnet action is
\be
S_{GB}= \gamma\int d^4 x\,\sqrt{-g} \left \{ R^2
- 4 R_{\mu\nu}R^{\mu\nu} + R_{\mu\nu\lambda\delta}
R^{\mu\nu\lambda\delta} \right \},  
\ee
given that one performs the following identification
\be
\tilde{\alpha} = \alpha - \gamma \hskip 1 cm \tilde{\beta} = \beta + 4\gamma
\ee
The theory defined by the action (apart from the topological term) 
(\ref{action2}) has been studied 
since the early days of the emergence of Einstein's theory of gravity.
\cite{treder,stelle,accioly,schmidt}.
We can borrow the non-relativistic potential from \cite{treder,schmidt} 
\be
V(r)= GM\left \{ -{1\over r} + {4\over 3}{e^{-m_1 r}\over r} - {1\over 3}
{e^{-m_0 r}\over r} \right \}  
\label{potential}
\ee
where~\footnote{
One thing to notice is that for finite masses the potential 
for large distances reduces to the usual Newtonian limit
and near the origin it is finite and we have
$V(r)= {G M\over 3} \left \{ (m_0 - 4 m_1) - (4 m_1^2 - m_0^2){ r\over 2} 
+O(r^2) \right \}$ } 
\be
m_0^2 = {1\over{( 3\tilde{\alpha} +\tilde{\beta})32\pi G}} \hskip 2 cm  
m_1^2 = -{1\over \tilde{\beta}16\pi G}
\label{masses}
\ee
There are three modes
in the theory one with a vanishing mass which
gives the Newton force and the two massive modes which create
Yukawa-type interactions.

One needs to impose the condition that there are no tachyons
in the theory which yields ~\footnote{The theory which results when one from the onset 
sets $3 \tilde{\alpha} = -\tilde{\beta}$
is known as Einstein-Bach-Weyl gravity and it was studied in \cite{treder}.
Incidentally there seems to be a discontinuity in the spirit of van-Dam-Veltman
and Zakharov; namely the non-relativistic limit and the limit of
$3 \tilde{\alpha} = -\tilde{\beta}$ do not seem to commute. 
We expect this puzzle will 
be resolved `a la Vainshtein but we shall not discuss it here.}

\be
3 \tilde{\alpha} +\tilde{\beta} > 0 \hskip 2 cm -\tilde{\beta} > 0 
\label{condition}
\ee

We have gotten rid off the tachyons but the second term
in the potential (\ref{potential}) has a wrong sign. 
Ghosts are expected for generic values of $\tilde{\alpha}$ and 
$\tilde{\beta}$.~\footnote{I am grateful for S. Deser for 
stressing this point. I. Kogan  
pointed out to me that negative norm states (`radions') appear also 
in the brane world theories with negative tension branes [See the
discussion and references in \cite{pilo}].}
Fortunately we can also get rid of the ghost by choosing 
$\tilde{\beta}\sim 0$. $m_1$ becomes infinitely heavy and decouples 
form the theory. One then has the following potential~\footnote{An other interesting theory is 
when $m_0= m_1$ which can be obtained by setting
$2\tilde{\alpha} = -\tilde{\beta}$ the non-relativistic potential reads 
$V(r)= GM\left \{ -{1\over r} + {e^{-m_0 r}\over r}  \right \}$
In the limit of large $\tilde{\alpha}$ 
the mass goes to zero ( $m_0 \rightarrow 0$) 
and the potential vanishes. The theory in a sense becomes topological  
and the propagating graviton disappears. 
The Lagrangian reduces to 
${\cal{L}}= \tilde{\alpha} \left \{ R^2
- 2 R_{\mu\nu}R^{\mu\nu}\right \}$,
or in terms of the starting action (\ref{action1}) whenever 
$ 2\alpha + \beta + 6\gamma = 0$ is satisfied one gets a theory with no
Newtonian limit for large  $\alpha$. Needless to say that this theory
is not compatible with the experiments. The massive modes should in fact 
be quite heavy to escape large violations of Newton's law in large 
distances.} 
\be
V(r)= -GM\left \{ {1\over r} + {1\over 3}
{e^{-m_0 r}\over r} \right \}  
\label{pot}
\ee

Now we have 
\be
m_0^2 = {1\over{ \tilde{\alpha}96\pi G}}
\ee

Recent experiments in gravity \cite{hoyle} tested $1/r^2$ law 
down to sub-millimeter $(0.1 mm)$.
One can safely take $m_0$ of the
order of, or bigger than, ${10/mm}$  
with no observable effects as far as gravitational potential is concerned.
We still need to check whether
the bending of light by the Sun's gravitational field predicted by the higher
curvature theory  (\ref{action2}) is consistent with the experiments or not.
This computation was carried out in \cite{accioly}.
Below I shall summarize their result. Around the Sun the metric reads
\be
ds^2 = g_{00} dt^2 - f(r) d\vec{x}^2
\ee
where
\be
&&g_{00}= 1+ {2M G\over r}\left 
\{ -1 -{1\over 3}e^{-m_0 r}+ {4\over 3}e^{-m_1 r} \right \} \nonumber \\ 
&& f(r)=1- {2M G\over r}\left 
\{ -1 +{1\over 3}e^{-m_0 r}+ {2\over 3}e^{-m_1 r} \right \}
\ee
Then for the null geodesic $ds^2 = 0 $ the index of refraction reads
\be
n(r) = \sqrt{{f\over g_{00}}} = 1+ {2M G\over r} -{2M G\over r}e^{-m_1 r} 
+O(G^2)
\ee
 
Crucial observation is that the massive scalar mode 
( $m_0$ ) do not contribute to the deflection of light. 
The contribution of the ghost state is exponentially
suppressed and goes to zero for infinite $(m_1)$ which
is the limit we consider here.

To conclude in this letter we have studied the various limits of higher 
curvature gravity in four dimensions. Higher curvature terms generate
massive modes in addition to the usual
massless  mode in gravity. In particular the Lagrangian
\be
{\cal{L}} = {R\over 16\pi G } + {\tilde{\alpha}} R^2
\ee
has a massless and a massive mode whose mass can be
$\geq 10^{-3}$eV without contradicting the experimental 
tests of gravity: the $1/r^2$ law and the bending of light by the Sun.

We need to stress a crucial point 
with regard to the experiments on large extra dimensions. 
Theories with large extra dimensions predict non-relativistic
potential of the form \cite{arkani,randall,sfetsos}
\be
V(r)= -GM\left \{{1\over r} + a{e^{-b r}\over r}  \right \}
\ee 
Clearly this potential is of the form one obtains in
higher curvature gravity (\ref{pot}) if $a$ is of the order of
unity. In fact starting from higher dimensions and compactifying on spheres 
or tori  $a$ turns out to be $O(1)$ or 
for Calabi-Yau compactifications $a$ can be at most 
as big as 20 \cite{sfetsos}. Therefore if $a$ is not too large~\footnote{ 
See \cite{dimopoulos} for the possibility of large $a$ in certain 
supersymmetric theories when
moduli fields and the dilaton are taken into account}
the non-relativistic potentials derived from large extra dimensions will
coincide with the potential derived from higher curvature
gravity. It would be hard to decide whether the experimentalists are measuring
the effects of large extra dimensions or the effects of massive modes in higher
curvature gravity; namely the effects of higher dimensional operators 
with {\it{large}} numerical coefficients.

\section{Addendum}
After the submission of the manuscript to the web I was informed that several related 
papers appeared before. In particular \cite{dva} and \cite{kiritsis} deal with $R^2$ 
terms and more. The work of \cite{dva} deserves special mention since it has 
a strong overlap with the present work but their emphasis is rather different. 
For the continuity/discontinuity arguments in the higher curvature theory,
I refer the reader to the following recent article \cite{neu}. 
I duly thank the authors of the above works for the correspondence.

\section{Acknowledgments}
I would like to thank S. Deser, A. Ibarra, I. Kogan, V. K. Onemli, 
G. Ross and R. Sturani for reading the manuscript and for useful discussions.  
This work is supported by  PPARC Grant PPA/G/O/1998/00567.

\vskip 1cm


\myend
\begin{thebibliography}{99}


\bibitem{arkani}
N.~Arkani-Hamed, S.~Dimopoulos and G.~Dvali,
Phys.\ Lett.\ B {\bf 429}, 263 (1998)
[hep-ph/9803315].

\bibitem{anton}
I.~Antoniadis, N.~Arkani-Hamed, S.~Dimopoulos and G.~Dvali,
Phys.\ Lett.\ B {\bf 436}, 257 (1998)
[hep-ph/9804398].




\bibitem{randall}
L.~Randall and R.~Sundrum,
Phys.\ Rev.\ Lett.\  {\bf 83}, 4690 (1999)
[hep-th/9906064].

\bibitem{ross}
I.~I.~Kogan, S.~Mouslopoulos, A.~Papazoglou, G.~G.~Ross and J.~Santiago,
Nucl.\ Phys.\ B {\bf 584}, 313 (2000)
[hep-ph/9912552].


\bibitem{gregory}
R.~Gregory, V.~A.~Rubakov and S.~M.~Sibiryakov,
Phys.\ Rev.\ Lett.\  {\bf 84}, 5928 (2000)
[hep-th/0002072].

\bibitem{dvali}
G.~Dvali, G.~Gabadadze and M.~Porrati,
Phys.\ Lett.\ B {\bf 484}, 112 (2000)
[hep-th/0002190].

\bibitem{veltman}
H.~van Dam and M.~Veltman,
Nucl.\ Phys.\ B {\bf 22}, 397 (1970).

\bibitem{zakharov}
V.I. Zakharov, JETP Lett. {\bf 12} 312 (1970)

\bibitem{deffayet}
C.~Deffayet, G.~Dvali, G.~Gabadadze and A.~Vainshtein,
[hep-th/0106001].


\bibitem{deser}
D.~G.~Boulware and S.~Deser,
Phys.\ Rev.\ D {\bf 6} (1972) 3368.

\bibitem{vainshtein}
A.~I.~Vainshtein,
Phys.\ Lett.\ B {\bf 39}, 393 (1972).


\bibitem{kogan}
I.~I.~Kogan, S.~Mouslopoulos and A.~Papazoglou,
Phys.\ Lett.\ B {\bf 503}, 173 (2001)
[hep-th/0011138].

\bibitem{porrati}
M.~Porrati,
Phys.\ Lett.\ B {\bf 498}, 92 (2001)
[hep-th/0011152].

\bibitem{higuchi}
A.~Higuchi,
Nucl.\ Phys.\ B {\bf 282}, 397 (1987);
Nucl.\ Phys.\ B {\bf 325} (1989) 745.

\bibitem{talk}
Talk presented by I. I. Kogan at the XXXVI Rencontres de Moriond, 
``Electroweak Interactions and Unified Theories'', Les Arcs.

\bibitem{groom}
D.~E.~Groom {\it et al.}  [Particle Data Group Collaboration],
Eur.\ Phys.\ J.\ C {\bf 15}, 1 (2000).

\bibitem{boulware}
D.~G.~Boulware and S.~Deser,
Phys.\ Rev.\ Lett.\  {\bf 55}, 2656 (1985).

\bibitem{zwiebach}
B.~Zwiebach,
Phys.\ Lett.\ B {\bf 156}, 315 (1985).

\bibitem{pilo}
I.~I.~Kogan, S.~Mouslopoulos, A.~Papazoglou and L.~Pilo,
[hep-th/0105255].



\bibitem{treder}
[The early references are in the following paper]
 H.~Von Borzeszkowski, H.~J.~Treder and W.~Yourgrau,
Annalen Phys.\  {\bf 35} 471 (1978)

\bibitem{stelle}
K.~S.~Stelle,
Gen.\ Rel.\ Grav.\  {\bf 9}, 353 (1978).

\bibitem{accioly}
A. Accioly, A. Azeredo,, H. Mukai and E.de R. Neto,
Prog.\ Theo.\ Phys {\bf104} 103 (2000)

\bibitem{schmidt}
H.~J.~Schmidt,
[gr-qc/0106037].


\bibitem{hoyle}
C.~D.~Hoyle, U.~Schmidt, B.~R.~Heckel, E.~G.~Adelberger, J.~H.~Gundlach, D.~J.~Kapner and H.~E.~Swanson,
Phys.\ Rev.\ Lett.\  {\bf 86}, 1418 (2001)
[hep-ph/0011014].


\bibitem{sfetsos}
A.~Kehagias and K.~Sfetsos,
Phys.\ Lett.\ B {\bf 472}, 39 (2000)
[hep-ph/9905417].


\bibitem{dimopoulos}
S.~Dimopoulos and G.~F.~Giudice,
Phys.\ Lett.\ B {\bf 379}, 105 (1996)
[hep-ph/9602350].


\bibitem{dva}
G.~R.~Dvali, G.~Gabadadze, M.~Kolanovic and F.~Nitti,
[hep-th/0106058].

\bibitem{kiritsis}
E.~Kiritsis, N.~Tetradis and T.~N.~Tomaras,
JHEP {\bf 0108}, 012 (2001)
[hep-th/0106050].

\bibitem{neu}
I.~P.~Neupane,
[hep-th/0108194].

\end{thebibliography}
